**Magnetic structure and ferroelectric polarization of MnWO$_4$ investigated by density functional calculations and classical spin analysis**

Chuan Tian[1], Changhoon Lee[1], Hongjun Xiang[2], Yuemei Zhang[1], Christophe Payen[3], Stéphane Jobic[3], and Myung-HwanWhangbo[1]*

[1] Department of Chemistry, North Carolina State University, Raleigh, NC 27695-8204

[2] National Renewable Energy Laboratory, Golden, Colorado 80401

[3] Institut des Matériaux Jean Rouxel, Université de Nantes, CNRS, 2 rue de la Houssinière, BP 32229, 44322 Nantes, France

**Abstract**

The ordered magnetic states of $MnWO_4$ at low temperatures were examined by evaluating the spin exchange interactions between the $Mn^{2+}$ ions of $MnWO_4$ on the basis of first principles density functional calculations and by performing classical spin analysis with the resulting spin exchange parameters. Our work shows that the spin exchange interactions are frustrated within each zigzag chain of $Mn^{2+}$ ions along the c-direction and between such chains of $Mn^{2+}$ ions along the a-direction. This explains the occurrence of a spiral-spin order along the c- and a-directions in the incommensurate magnetic state AF2, and that of a ↑↑↓↓ spin order along the c- and a-directions in the commensurate magnetic state AF1. The ferroelectric polarization of $MnWO_4$ in the spiral-spin state AF2 was examined by performing Berry phase calculations for a model superstructure to find that the ferroelectric polarization occurs along the b-direction, in agreement with experiment.

# I. Introduction

Manganese tungstate $MnWO_4$ is made up of $MnO_6$ octahedra with high-spin $Mn^{2+}$ ($d^5$) ions and $WO_6$ octahedra with diamagnetic $W^{6+}$ ($d^0$) ions [1-3]. The $MnO_6$ octahedra share their *cis* edges to form zigzag $MnO_4$ chains along the c-direction (**Fig. 1a**). Similarly, the $WO_6$ octahedra share their *cis* edges to form zigzag $WO_4$ chains along the c-direction (**Fig. 1b**). These $MnO_4$ and $WO_4$ chains share their octahedral corners to form the three-dimensional structure of $MnWO_4$ (**Fig. 1c**). As a result, layers of $Mn^{2+}$ ions parallel to the bc-plane (hereafter, the //bc-layers of $Mn^{2+}$ ions) alternate with layers of $W^{6+}$ ions parallel to the bc-plane along the a-direction. The neutron diffraction study by Lautenschläger *et al.* [3] established that $MnWO_4$ undergoes three magnetic phase transitions below 14 K, namely, the paramagnetic to the AF3 state at $T_N$ (13.5 K), the AF3 to the AF2 state at $T_2$ (12.3 K), and the AF2 to the AF1 state at $T_1$ (8.0 K). The magnetic structures of the AF3 and AF2 states are incommensurate with the propagation vector (-0.214, 0.5, 0.457), while that of the AF1 state is commensurate with the propagation vector (-0.25, 0.5, 0.5). Ehrenberg *et al.* [4] analyzed the spin wave dispersion curves of the magnetic state AF1 of $MnWO_4$, determined from inelastic neutron scattering measurements [4], in terms of nine spin exchange parameters (**Fig. 2**, **Table 1**); the exchanges $J_1$ and $J_2$ in the zigzag chains of $Mn^{2+}$ ions along the c-direction (hereafter, the //c-chains of $Mn^{2+}$ ions), the exchanges $J_3$ and $J_4$ between adjacent //c-chains of $Mn^{2+}$ ions in each //bc-layer of $Mn^{2+}$ ions, and the exchanges $J_5 - J_9$ between adjacent //bc-layers of $Mn^{2+}$ ions along the a-direction. So far, it has not been tested whether the values of $J_1 - J_9$ extracted by Ehrenberg *et al.* [4] are consistent with other magnetic properties and the electronic structure of $MnWO_4$.

Recently, MnWO$_4$ has received much attention due to the finding that it exhibits ferroelectric (FE) polarization in the AF2 state [5-7], because this state has a spiral-spin order [6,7] and hence has no inversion symmetry [8-10]. It was found that the spin spiral of the AF2 state propagates along the //c-chain direction as well as along the interlayer direction (i.e., the a-direction) [7]. This implies that MnWO$_4$ has spin frustration in the exchange interactions within each //c-chain and between the //c-chains along the a-direction, because noncollinear spin arrangements occur generally to reduce the extent of geometric spin frustration [11-15]. Indeed, MnWO$_4$ has been known to be a moderately spin frustrated system with the frustration parameter is f = $-\theta/T_N \approx 5$, where the Curie-Weiss temperature $\theta$ is approximately $-75$ K and the Néel temperature $T_N$ is 13.5 K [5,16]. In the spin exchange values of Ehrenberg *et al.* (Table 1) [4], the nearest-neighbor intrachain exchange $J_1$ is antiferromagnetic (AFM) while the next-nearest-neighbor intrachain exchange $J_2$ is ferromagnetic (FM). The latter means that the spin exchanges within each //c-chain are not spin-frustrated, which is inconsistent with the occurrence of spiral-spin within each //c-chain [7]. The incommensurate (AF2 and AF3) and the commensurate (AF1) states have a common feature in their propagation vectors, i.e., the magnetic order along the b-direction is AFM. The reason for this commensurate component is unclear, although one might speculate if it is a consequence of spin exchange interactions other than $J_3$ and $J_4$ because the latter two are quite weak according to Ehrenberg *et al.* [4]. One drawback of spin exchange parameters deduced from experiment is that the physical data of a given magnetic system (e.g., magnetic susceptibility, inelastic neutron scattering and heat capacity data) are often fitted equally well by more than one set of exchange parameters, as found, for example, for (VO)$_2$P$_2$O$_7$ [17,18], $Na_3Cu_2SbO_6$ and $Na_2Cu_2TeO_6$ [19-23], and $Bi_4Cu_3V_2O_{14}$ [24-27]. Ultimately, the correct set of exchange parameters should be consistent with the electronic structure of a magnetic system under consideration because it is the electronic structure that determines the magnetic energy spectrum [28-31]. In view of the fact that the magnetoelectric properties of $MnWO_4$ have attracted much attention in recent years, it is important to investigate the spin exchange interactions and the FE polarization of $MnWO_4$ in terms of electronic structure calculations.

In the present work, we examine the magnetic and FE properties of $MnWO_4$ on the basis of first principles density functional theory (DFT) electronic structure calculations. We first evaluate the nine spin exchange parameters of $MnWO_4$ by carrying out mapping analysis based on DFT calculations [29]. Then, we perform classical spin analysis [30] with the resulting spin exchange parameters to probe the incommensurate magnetic structure of $MnWO_4$. Finally, we evaluate the FE polarization of $MnWO_4$ by using the Berry phase method [32,33].

## II. Computational details

Our calculations employed the Vienna ab-initio simulation package [34-36], the generalized gradient approximations (GGA) for the exchange and correlation corrections [37], the plane-wave cutoff energy of 400 eV, 196 k-points for the irreducible Brillouin zone, and the threshold of $10^{-6}$ eV for the self-consistent-field convergence of the total electronic energy. To properly describe the electron correlation of the Mn 3d states, the GGA plus on-site repulsion U (GGA+U) method [38] was employed with an effective U = 4 and 6 eV on the Mn atom. The nine spin exchange parameters were evaluated by performing GGA+U calculations for a number of ordered spin states of $MnWO_4$ (see below). On the basis of the resulting spin exchange parameters, we examined the magnetic structure of $MnWO_4$ by performing classical spin analysis as described elsewhere [30]. For the ferroelectricity driven by a magnetic order, it is essential to take into consideration spin-orbit coupling (SOC) effects in electronic structure calculations [13-15]. Thus, to estimate the FE polarization of $MnWO_4$ in the AF2 state, we performed GGA+U+SOC calculations for a model spiral-spin state of $MnWO_4$ (see below), which is designed to simulate the AF2 state of $MnWO_4$. Then, the FE polarization was calculated by using the Berry phase method [32,33].

### III. Spin exchange parameters

In the magnetic state AF1 the ↑↑↓↓ spin order occurs along the c- and the a-directions, and the ↑↓↑↓ spin arrangement along the b-direction. In addition, the phase of the ↑↑↓↓ arrangement along the a-direction is shifted such that four different ↑↑↓↓ arrangements occur (i.e., ↑↓↓↑, ↓↓↑↑, ↓↑↑↓, ↑↑↓↓). Consequently, the propagation vector of the AF1 state becomes (-0.25, 0.5, 0.5). In general, the ↑↑↓↓ arrangement along a certain direction occurs when there is geometric spin frustration, as found for the $CuO_2$ ribbon chains of $LiCuVO_4$ and $LiCu_2O_2$ [13]. The propagation vector of the incommensurate states AF2 and AF3, (-0.214, 0.5, 0.457), is slightly different from that of the AF1 state. To gain insight into the occurrence of the ordered magnetic states AF1 and AF2 of $MnWO_4$, we evaluate its spin exchange parameters $J_1 - J_9$ and discuss their trends.

## A. Mapping analysis

To evaluate $J_1 - J_9$, we examine the 10 ordered spin states defined in **Figs. 3-5**. The relative energies of these states determined from our GGA+U calculations are summarized in **Table 2**. To extract the values of the spin exchange parameters $J_1 - J_9$, we express the total spin exchange interaction energies of the 10 ordered spin states in terms of the spin Hamiltonian,

$$\hat{H} = -\sum_{i<j} J_{ij} \hat{S}_i \cdot \hat{S}_j, \qquad (1)$$

where $J_{ij}$ ($= J_1 - J_9$) is the spin exchange parameter for the spin exchange interaction between the spin sites i and j, while $\hat{S}_i$ and $\hat{S}_j$ are the spin angular momentum operators at the spin sites i and j, respectively. Then, by applying the energy expressions obtained for spin dimers with N unpaired spins per spin site (in the present case, N = 5) [39,40], the total spin exchange energies of the 10 ordered spin states (per two formula units) are written as

$$E_{FM} = (-2J_1 - 2J_2 - 2J_3 - 2J_4 - 2J_5 - 2J_6 - 2J_7 - 2J_8 - 2J_9)(N^2/4)$$

$$E_{AF1} = (+2J_1 - 2J_2 + 2J_3 - 2J_4 - 2J_5 + 2J_6 + 2J_7 + 2J_8 + 2J_9)(N^2/4)$$

$$E_{AF2} = (+2J_2 - 2J_4 - 2J_5)(N^2/4)$$

$$E_{AF3} = (-2J_1 - 2J_2 + 2J_3 + 2J_4 - 2J_5 + 2J_6 + 2J_7 - 2J_8 - 2J_9)(N^2/4)$$

$$E_{AF4} = (-2J_1 - 2J_2 - 2J_3 - 2J_4 + 2J_5 + 2J_6 + 2J_7 + 2J_8 + 2J_9)(N^2/4)$$

$$E_{AF5} = (-3J_1 - 3J_2 - J_3 - J_4 - J_5 - J_6 - J_7 - J_8 - J_9)(N^2/8)$$

$$E_{AF6} = (-3J_1 - 3J_2 - 3J_3 - 3J_4 - 3J_5 - 3.5J_6 - 3J_7 - 3J_8 - 3J_9)(N^2/8)$$

$$E_{AF7} = (-3J_1 - 3J_2 - 3J_3 - 3J_4 - 3J_5 - 3J_6 - 3J_7 - 3.5J_8 - 3J_9)(N^2/8)$$

$$E_{AF8} = (-3J_1 - 3J_2 - 3J_3 - 3J_4 - 3J_5 - 3.5J_6 - 3.5J_7 - 3J_8 - 3J_9)(N^2/8)$$

$$E_{AF9} = (-3J_1 - 3J_2 - 3J_3 - 3J_4 - 3J_5 - 3J_6 - 3J_7 - 3J_8 - 3.5J_9)(N^2/8) \quad (2)$$

Thus, by mapping the relative energies of the 10 ordered spin states determined by the GGA+U calculations onto the corresponding relative energies determined from the above spin exchange energies, we obtain the values of $J_1 - J_9$ summarized in **Table 1**.

### B. Trends in the spin exchange interactions

Let us first comment on the values of $J_1 - J_9$ calculated in the previous section. In the mean field theory [41], which is valid in the high-temperature paramagnetic limit, the Curie-Weiss temperature $\theta$ is related to the spin exchange parameters of MnWO$_4$ as follows:

$$\theta = \frac{S(S+1)}{3k_B} \sum_i z_i J_i, \quad (3)$$

where the summation runs over all nearest neighbors of a given spin site, $z_i$ is the number of nearest neighbors connected by the spin exchange parameter $J_i$, and S is the spin quantum number of each spin site Mn$^{2+}$ (i.e., S = 5/2 in the present case). Thus, according to the spin exchange paths defined in **Fig. 2**, $\theta$ is expressed as

$$\theta = \frac{20(J_1 + J_2 + J_3 + J_4 + J_5 + J_6 + J_7 + J_8 + J_9)}{k_B} \quad (4)$$

From the $J_i$ values from the GGA+U calculations (**Table 1**), the Curie-Weiss temperature is found to be $\theta_{cal} = -357$ and $-232$ K for U = 4 and 6 eV, respectively. The experimental Curie-Weiss temperature is $\theta_{exp} \approx -75$ K, so that $\theta_{cal}$ is greater than $\theta_{exp}$ by a factor of approximately 3 – 5. The latter means that the calculated $J_i$ values are overestimated by a factor of approximately 3 – 5, which is consistent with the finding that DFT electronic structure calculations generally overestimate the magnitude of spin exchange interactions by a factor of approximately 4 [39,42-44]. In terms of the $J_i$ values of Ehrenberg *et al.* [4] (**Table 1**) the Curie-Weiss temperature is calculated to be −32 K, so that their $J_i$ values are underestimated by a factor of approximately 2.5 as far as the Curie-Weiss temperature is concerned.

Our calculations (**Table 1**) show that both $J_1$ and $J_2$ are AFM, and $J_2/J_1 \gg 0.25$, so that geometric spin frustration exists within each //c-chain. This finding accounts for the occurrence, in each //c-chain, of the spiral-spin order in the AF2 state and the ↑↑↓↓ spin order in the AF1 state. To see if there exists spin frustration along the a-direction, we consider four commensurate spin arrangements of MnWO$_4$ generated by using the //c-chains with ↑↑↓↓ spin order as the building units (**Figure 6**). In the AF1 state (**Fig. 6a**), the //c-chains have a ↑↓↑↓ order (i.e., an AFM coupling) along the b-direction and a ↑↑↓↓ spin order along the a-direction. The AF4 state (**Fig. 6b**) differs from the AF1 state, only in that the //c-chains have a ↑↓↑↓ order along the a-direction. The AF5 (**Fig. 6c**) and AF6 (**Fig. 6d**) states differ from the AF1 and AF4 states only in that the //c-chains have a ↑↑↑↑ order (i.e., an FM coupling) along the b-direction, respectively. Per spin site, these four commensurate spin arrangements give rise to the interchain spin exchange energies listed in **Table 3**. The exchange parameters $J_1 – J_9$ of Ehrenberg *et al.* as well as those obtained from our GGA+U calculations predict that the AF1 state is the most stable state of the four. The $J_1 – J_9$ values from the GGA+U calculations show that the //c-chains with ↑↑↓↓ spin order prefer to have a ↑↓↑↓ order rather than a ↑↑↑↑ order along the b-direction, in agreement with experiment. Furthermore, the $J_1 – J_9$ values from the GGA+U calculations predict that the AF4 state (i.e., a ↑↓↑↓ order along the a-direction) is close in energy to the AF1 state (i.e., a ↑↑↓↓ order along the a-direction). In other words, along the a-direction, the interaction between the //c-chains with the ↑↑↓↓ spin order is effectively frustrated because the spin order can be either ↑↑↓↓ or ↑↓↑↓, the extent of which may be reduced by an incommensurate spiral-spin order along the a-direction. This topic will be discussed further in the next section.

**IV. Classical spin analysis**

To examine the occurrence of the incommensurate magnetic structure (i.e., the AF2 and AF3 states) in MnWO$_4$, we calculate the total spin exchange energy of MnWO$_4$ by using the Freiser method [30,45]. This approach assumes that spins adopt all possible directions in space (i.e., the classical spin approximation), and the spin exchange interactions are isotropic (i.e., a Heisenberg description). These assumptions are appropriate for MnWO$_4$, because the local magnetic anisotropy of the high spin Mn$^{2+}$ (S = 5/2, L = 0) ions is very small. Indeed, MnWO$_4$ is well described as a Heisenberg antiferromagnet, as shown by the very small anisotropy of the paramagnetic susceptibility [5]. This also means that the complex low-temperature magnetic properties of MnWO$_4$ do not arise from a competition between the local anisotropy and the spin exchange interactions, but from the frustration of the spin exchange interactions.

In a long-range ordered magnetic state of a magnetic system, the spin sites μ (= 1, 2, ... , m) of its unit cell located at the coordinate origin (i.e., the lattice vector **R** = 0) have the spin moments $\sigma_\mu^0$. For a magnetic solid with repeat vectors **a**, **b** and **c**, the ordered spin arrangement is described by the spin functions $\sigma_\mu(\mathbf{k})$,

$$\sigma_\mu(\mathbf{k}) = \frac{1}{\sqrt{M}} \sum_\mathbf{R} \sigma_\mu^0 \exp(i\mathbf{k} \cdot \mathbf{R}), \tag{5}$$

where M is the number of unit cells in the magnetic solid, $\mathbf{k}$ is the wave vector, and $\mathbf{R}$ is the direct lattice vector [46]. The ordered magnetic state $\psi_i(\mathbf{k})$ (i = 1 – m) is then expressed as a linear combination of the spin functions $\sigma_\mu(\mathbf{k})$,

$$\psi_i(\mathbf{k}) = C_{1i}(\mathbf{k})\sigma_1(\mathbf{k}) + C_{2i}(\mathbf{k})\sigma_2(\mathbf{k}) + \ldots + C_{mi}(\mathbf{k})\sigma_m(\mathbf{k}). \tag{6}$$

To determine the energy $E_i(\mathbf{k})$ of the state $\psi_i(\mathbf{k})$ and the coefficients $C_{\mu i}(\mathbf{k})$ ($\mu$ = 1 – m), one needs to evaluate the spin exchange interaction energies $\xi_{\mu\nu}(\mathbf{k})$ between every two spin functions $\sigma_\mu(\mathbf{k})$ and $\sigma_\nu(\mathbf{k})$,

$$\xi_{\mu\nu}(\mathbf{k}) = -\sum_\mathbf{R} J_{\mu\nu}(\mathbf{R})\exp(i\mathbf{k} \cdot \mathbf{R}), \tag{7}$$

where $J_{\mu\nu}(\mathbf{R})$ = $J_1$, $J_2$, $J_3$, $J_4$, $J_5$, $J_6$, $J_7$, $J_8$, or $J_9$. The resulting interaction matrix $\Xi(\mathbf{k})$ is given by

$$\Xi(\mathbf{k}) = \begin{pmatrix} \xi_{11}(\mathbf{k}) & \xi_{12}(\mathbf{k}) & \ldots & \xi_{1m}(\mathbf{k}) \\ \xi_{21}(\mathbf{k}) & \xi_{22}(\mathbf{k}) & \ldots & \xi_{2m}(\mathbf{k}) \\ \ldots & \ldots & \ldots & \ldots \\ \xi_{m1}(\mathbf{k}) & \xi_{m2}(\mathbf{k}) & \ldots & \xi_{mm}(\mathbf{k}) \end{pmatrix}. \tag{8}$$

We obtain $E_i(\mathbf{k})$ of the state $\psi_i(\mathbf{k})$ by diagonalizing this matrix. This method predicts the superstructure of a magnetic system by finding the wave vector at which its global energy minimum occurs [30,45].

There are two $Mn^{2+}$ ion sites in a crystallographic unit cell of $MnWO_4$ (**Table 3a**), so that m = 2 in Eq. 8. The spin exchanges $J_1$ – $J_9$ occur for various pairs of spin sites ($\mu,\nu$) within a unit cell located at [0, 0, 0] as well as between adjacent unit cells [0, 0, 0] and [n, k, l] (n, k, l = −1, 0, +1), as summarized in **Table 3b**. Consequently, we obtain the following matrix elements $\xi_{\mu\nu}(\mathbf{k})$ ($\mu, \nu = 1, 2$),

$$\begin{aligned}
\xi_{11}(\mathbf{k}) = \xi_{22}(\mathbf{k}) = &-J_2[\exp(-i2\pi x_c) + \exp(+i2\pi x_c)] \\
&- J_4[\exp(-i2\pi x_b) + \exp(+i2\pi x_b)] \\
&- J_5[\exp(-i2\pi x_a) + \exp(+i2\pi x_a)]
\end{aligned}$$

$$\begin{aligned}
\xi_{12}(\mathbf{k}) = \xi_{21}(\mathbf{k})^* = &-J_1\{1 + \exp[i2\pi(-x_c)]\} \\
&- J_3\{\exp[i2\pi(+x_b)] + \exp[i2\pi(+x_b - x_c)]\} \\
&- J_6\{\exp[i2\pi(+x_a + x_b - x_c)] + \exp[i2\pi(-x_a + x_b)]\} \\
&- J_7\{\exp[i2\pi(-x_a + x_b - x_c)] + \exp[i2\pi(+x_a + x_b)]\} \\
&- J_8\{\exp[i2\pi(+x_a - x_c)] + \exp[i2\pi(-x_a)]\} \\
&- J_9\{\exp[i2\pi(+x_a)] + \exp[i2\pi(-x_a - x_c)]\}
\end{aligned} \quad (9)$$

where $x_a$, $x_b$ and $x_c$ are dimensionless numbers [46].

Thus, at any given wave vector **k**, one can determine the numerical values of $\xi_{\mu\nu}(\mathbf{k})$ ($\mu, \nu = 1, 2$) by using the $J_1 - J_9$ values listed in **Table 1** and hence diagonalize the $\Xi(\mathbf{k})$ matrix to obtain $E_i(\mathbf{k})$ (i = 1, 2). Then, the propagation vector **q** of the incommensurate magnetic state is determined as the **k** value at which the lower energy $E_1(\mathbf{k})$ has the minimum. Our calculations show that **q** = (−0.29, 0.5, 0.44) from the $J_1 - J_9$ values of Ehrenberg *et al.*, and (−0.35, 0.5, 0.49) and (−0.36, 0.5, 0.48) from the calculated $J_1 - J_9$ values with U = 4 and 6 eV, respectively. These results are in qualitative agreement with the appearance of the incommensurate AF2 state with **q** = (−0.214, 0.5, 0.457). From this finding and our discussion in the previous section (**Fig. 6**), it is clear that the AFM coupling along the b-direction both in the incommensurate state AF2 and in the commensurate state AF1 arises to lower the energy associated with the spin exchange interactions other than the weak interchain exchanges $J_3$ and $J_4$.

The observation of the collinear commensurate state AF1 as the magnetic ground state below 8 K is due most likely to a weak local magnetic anisotropy of $Mn^{2+}$, which is not included in the Heisenberg model. Since the entropy is greater in the spiral-spin incommensurate state AF2 than in the collinear commensurate AF1 state, the AF2 state would be energetically favored over the AF1 state at temperature higher than 8 K. It is interesting to note that doping the $Mn^{2+}$ sites of $MnWO_4$ with a small amount of $Fe^{2+}$ ($d^6$, S = 2) ions stabilizes the AF1 state [47], which is due to the large local magnetic anisotropy of the high-spin $Fe^{2+}$ ions.

## V. Ferroelectric polarization

In this section, we examine the FE polarization of the spiral-spin state AF2 of $MnWO_4$. The spin-spiral plane is perpendicular to the ac-plane and is tilted away from the a-axis by 35° (**Fig. 7**). The incommensurate propagation vector of the AF2 state is (-0.214, 0.5, 0.457), the closest commensurate approximation of which is (-0.25, 0.5, 0.5). The latter requires the use of the supercell (4a, 2b, 2c) for our GGA+U+SOC calculations of FE polarization. **Fig. 8** shows the AFM and FM arrangements, along the b-direction, of the //c-chains with spiral-spin order. In both the AFM and the FM arrangements, each //c-chain has the same chirality of spin spiral, and hence leads to a same sign of FE polarization. In addition, the interchain exchange interactions $J_3$ and $J_4$ are very weak, as already mentioned. Thus, for the purpose of FE polarization, we assume the FM ordering of the //c-chains along the b-direction and hence employ the supercell (4a, b, 2c) for our calculations. In making the spiral-spin arrangement with the (4a, b, 2c) supercell, we employed the plane of the spin spiral defined in **Fig. 7** so that the spins spiral along the a-and c-directions. In our GGA+U+SOC calculations for the electronic structure of this model spiral-spin state, the atom positions of the (4a, b, 2c) supercell were not relaxed. Our subsequent Berry phase calculations show that the FE polarization is along the positive b-direction with $P_b = 17.2$ μC/m². Experimentally, $P_b$ is found to be smaller than ~50 μC/m² [5].

It is of interest to examine the calculated FE polarization from the viewpoint of the Katsura-Nagaosa-Balatsky (KNB) model [48], which predicts that the FE polarization $\vec{P}$ of a spiral-spin chain with spins $\vec{S}_i$ and $\vec{S}_j$ at the adjacent spin sites connected by the vector $\vec{e}_{ij}$ is given by

$$\vec{P} \propto \vec{e}_{ij} \times (\vec{S}_i \times \vec{S}_j), \qquad (10)$$

according to which FE polarization arises only from spin-spiral chains of cycloidal type (i.e., those chains whose spin-spiral planes contain the chains). In the (4a, b, 2c) superstructure with the (-0.25, 0, 0.5) spiral-spin order, the spiral-spin propagation along the a-direction has a cyclodal component when the spins are projected on the ab-plane (**Fig. 10a**), while that along the c-direction has a cycloidal component when the spins are projected on the bc-plane (**Fig. 10b**). According to the KNB model, the cycloidal component in the ab-plane gives a positive FE polarization along the b-direction (**Fig. 10a**), whereas that in the bc-plane gives a negative FE polarization along the b-direction (**Fig. 10b**). Since the angle between the ab- and the spin-spiral planes is 35°, the FE polarization arising from the ab-plane cycloidal component is greater than that from the bc-plane cycloidal component. As a consequence, the net FE polarization is along the positive b-direction.

## VI. Concluding remarks

The spin exchange interactions of MnWO$_4$ extracted from the present GGA+U calculations reveal that the spin exchange interactions are frustrated within each //c-chain of Mn$^{2+}$ ions and between such //c-chains along the a-direction. This finding is in agreement with the experimental observation that a spiral-spin propagates along the c- and the a-directions in the incommensurate state AF2, and a ↑↑↓↓ spin arrangement occurs along the c- and a-directions in the collinear magnetic state AF1. The classical spin analysis with the extracted spin exchange parameters leads to an incommensurate state with propagation vector in qualitative agreement with that found for the AF2 state. The AFM coupling between the //c-chains along the b-direction does not result from the weak interchain exchanges $J_3$ and $J_4$ but from the combined effect of other strong spin exchange interactions. The Berry phase calculations for a model (4a, b, 2c) superstructure with spiral-spin order show FE polarization along the b-direction, in agreement with experiment.

**Acknowledgments**

The research was supported by the Office of Basic Energy Sciences, Division of Materials Sciences, U.S. Department of Energy, under Grant No. DE-FG02- 86ER45259, and by the NERSC Center (under Contract No. DE-AC02-05CH11231) and the HPC Center of the NCSU campus.

Table 1. Mn…Mn distances (in Å) associated with the spin exchange paths $J_1 - J_9$ of MnWO$_4$ and the values of $J_1 - J_9$ (in k$_B$K) determined by Ehrenberg *et al*. [4] from their neutron scattering study and by the present GGA+U calculations.

|  | Mn…Mn | Ehrenberg *et al*. | GGA+U (U = 4 eV) | GGA+U (U = 6 eV) |
|---|---|---|---|---|
| $J_1$ | 3.283 | -0.195 | -2.343 | -1.856 |
| $J_2$ | 4.992 | +0.414 | -4.222 | -2.691 |
| $J_3$ | 4.398 | -0.135 | -0.174 | -0.186 |
| $J_4$ | 5.753 | +0.021 | -0.418 | -0.209 |
| $J_5$ | 4.823 | -0.423 | -2.714 | -1.775 |
| $J_6$ | 6.561 | -1.273 | -2.378 | -1.334 |
| $J_7$ | 6.492 | +0.491 | -0.638 | -0.360 |
| $J_8$ | 5.873 | -0.509 | -3.364 | -2.146 |
| $J_9$ | 5.795 | +0.023 | -1.601 | -1.032 |

Table 2. Relative energies (in meV per two formula units) of the ordered spin states of MnWO$_4$ determined by the present GGA+U calculations.

|     | U = 4 eV | U = 6 eV |
| --- | --- | --- |
| FM  | 0.00   | 0.00   |
| A1  | -22.62 | -14.88 |
| A2  | -20.42 | -13.24 |
| A3  | -7.75  | -4.49  |
| A4  | -23.04 | -14.32 |
| A5  | -10.88 | -6.91  |
| A6  | -4.72  | -3.07  |
| A7  | -4.49  | -2.94  |
| A8  | -4.35  | -2.83  |
| A9  | -4.59  | -2.98  |

Table 3. Four arrangements of the //c-chains with ↑↑↓↓ spin order and their interchain spin exchange energies, E, per spin site.

|      | // b  | // a  | E (per Mn)                    | E ($k_B$K per Mn) [a] |         |         |
|------|-------|-------|-------------------------------|---------|---------|---------|
|      |       |       |                               | Case A  | Case B  | Case C  |
| AF1  | ↑↓↑↓  | ↑↑↓↓  | $+J_6 - J_7 + J_8 - J_9 + J_4$ | -2.75   | -3.92   | -2.30   |
| AF4  | ↑↓↑↓  | ↑↓↑↓  | $+J_5 + J_4$                  | -0.40   | -3.13   | -1.98   |
| AF5  | ↑↑↑↑  | ↑↑↓↓  | $-J_6 + J_7 + J_8 - J_9 - J_4$ | +1.21   | +0.40   | +0.07   |
| AF6  | ↑↑↑↑  | ↑↓↑↓  | $+J_5 - J_4$                  | −0.44   | −2.30   | −1.57   |

[a] Case A: Calculated from the $J_1 - J_9$ by Ehrenberg *et al*.

Case B: Calculated from the $J_1 - J_9$ of the present GGA+U calculations with U = 4 eV.

Case C: Calculated from the $J_1 - J_9$ of the present GGA+U calculations with U = 6 eV.

Table 4.    (a) Fractional coordinates of the spin sites in MnWO$_4$

| Spin site | Mn | x | y | z |
|---|---|---|---|---|
| 1 | Mn(1) | 0.5 | 0.6853 | 0.25 |
| 2 | Mn(2) | 0.5 | 0.3147 | 0.75 |

(b) Pairs (μ,ν) of the spin sites (μ, ν = 1, 2) leading to the spin exchanges $J_1 - J_9$ in MnWO$_4$ within a unit cell at [0, 0, 0] as well as between unit cells [0, 0, 0] and [n, k, l]

___

| Path | Within | Between | [n, k, l] |
|---|---|---|---|
| $J_1$ | (1,2) | (1,2) | [0, 0, -1] |
|  | (2,1) | (2,1) | [0, 0, 1] |
| $J_2$ |  | (1,1) | [0, 0, -1] & [0, 0, 1] |
|  |  | (2,2) | [0, 0, -1] & [0, 0, 1] |
| $J_3$ |  | (1,2) | [0, 1, 0] & [0, 1, -1] |
|  |  | (2,1) | [0, -1, 0] & [0, -1, 1] |
| $J_4$ |  | (1,1) | [0, 1, 0] & [0, -1, 0] |
|  |  | (2,2) | [0, 1, 0] & [0, -1, 0] |
| $J_5$ |  | (1,1) | [1, 0, 0] & [-1, 0, 0] |
|  |  | (2,2) | [1, 0, 0] & [-1, 0, 0] |
| $J_6$ |  | (1,2) | [1, 1, -1] & [-1, 1, 0] |
|  |  | (2,1) | [-1, -1, 1] & [1, -1, 0] |
| $J_7$ |  | (1,2) | [-1, 1-1] & [1, 1, 0] |
|  |  | (2,1) | [1, -1, 1] & [-1, -1, 0] |
| $J_8$ |  | (1,2) | [1, 0, -1] & [-1, 0, 0] |
|  |  | (2,1) | [-1, 0, 1] & [1, 0, 0] |
| $J_9$ |  | (1,2) | [1, 0, 0] & [-1, 0, -1] |
|  |  | (2,1) | [-1, 0, 0] & [1, 0, 1] |

___

**Figure captions**

Figure 1.    Perspective views of (a) a zigzag $MnO_4$ chain, (b) a zigzag $WO_4$ chain, and (c) the three-dimensional arrangement of the $MnO_4$ and $WO_4$ chains in $MnWO_4$. The Mn, W and O atoms are represented by large, medium and small spheres, respectively.

Figure 2.    (a) Four spin exchange paths $J_1 – J_4$ in $MnWO_4$ within each //bc-layer of $Mn^{2+}$ ions. (b) Five spin exchange paths $J_5 – J_9$ between adjacent //bc-layers of $Mn^{2+}$ ions in $MnWO_4$. The numbers 1 – 9 refer to the spin exchange paths $J_1 – J_9$, respectively.

Figure 3.    Ordered spin arrangements in each //bc-layer of $Mn^{2+}$ ions in the FM, A1, A2 and A3 states of $MnWO_4$.

Figure 4.    Ordered spin arrangements in two successive //bc-layers of $Mn^{2+}$ ions in the A4 and A5 states of $MnWO_4$.

Figure 5.    Ordered spin arrangements in four successive //bc-layers of $Mn^{2+}$ ions in the A6, A7, A8 and A9 states of $MnWO_4$.

Figure 6.    Spin arrangements of the (a) AF1, (b) AF4, (c) AF5 and (d) AF4 states of $MnWO_4$, which are generated in terms of the //c-chains with the ↑↑↓↓ spin order.

Figure 7.    Plane of spin spiral found in MnWO$_4$.

Figure 8.    Two arrangements of the //c-chains of Mn$^{2+}$ ions with spiral-spin order along the b-direction: (a) AFM and (b) FM. The zigzag chains are represented as straight chains to emphasize the spin spiral order.

Figure 9.    Perspective view of the commensurate spiral-spin arrangement of MnWO$_4$ within a (4a, b, 2c) supercell, which was employed for the GGA+U+SOC calculation of FE polarization. The plane of the spin spiral is defined as in Fig. 7, so that the spins spiral along the a- and c-directions.

Figure 10.   Cycloidal components of the spiral-spin state with the (4a, b, 2c) superstructure used to simulate the spiral-spin ordered state AF2. (a) The projection view of the spiral-spin propagation along the a-direction on the ab-plane. (b) that along the c-direction on the bc-plane, where the zigzag chain was represented as a straight chain to emphasize the spin spiral order. The polarization direction is represented by red arrows.

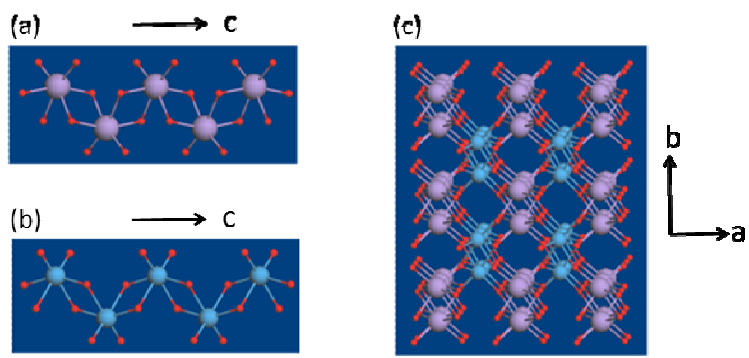

Figure 1.

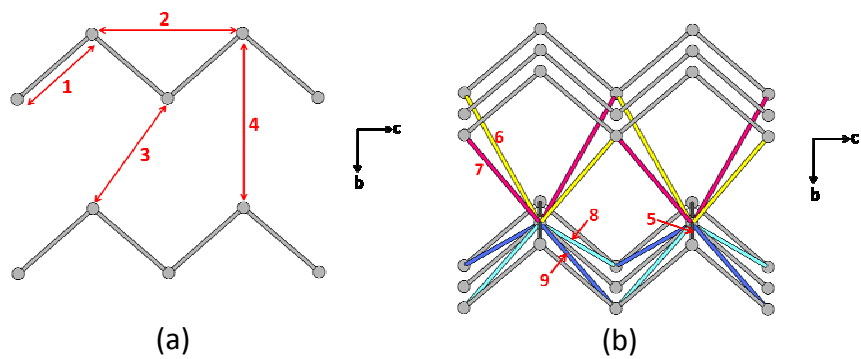

Figure 2.

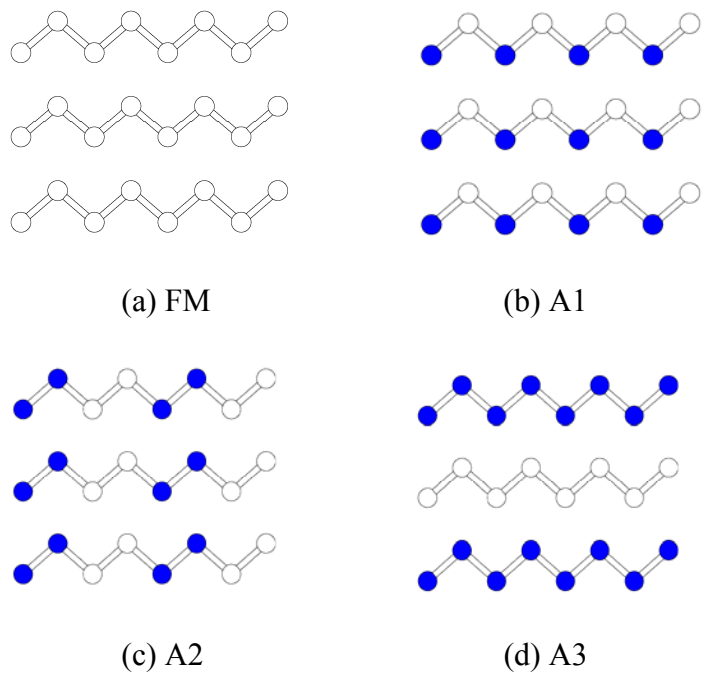

(a) FM  (b) A1

(c) A2  (d) A3

Figure 3.

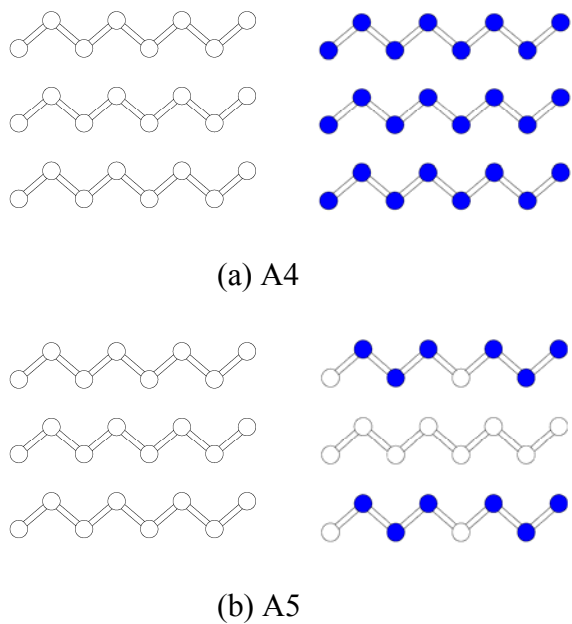

(a) A4

(b) A5

Figure 4.

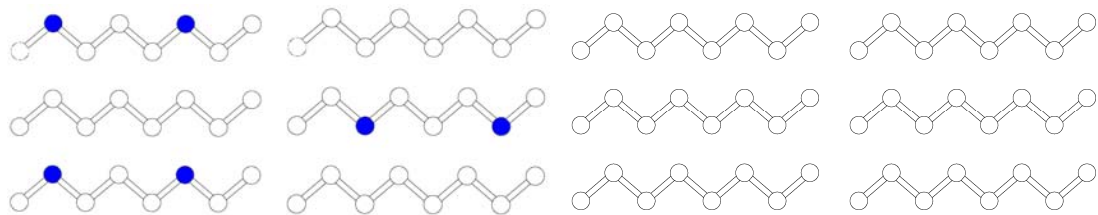

(a) A6

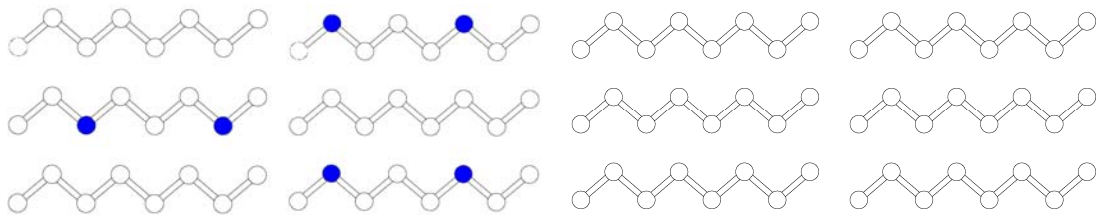

(b) A7

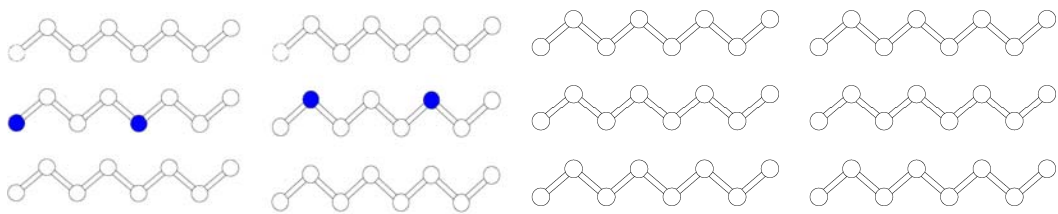

(c) A8

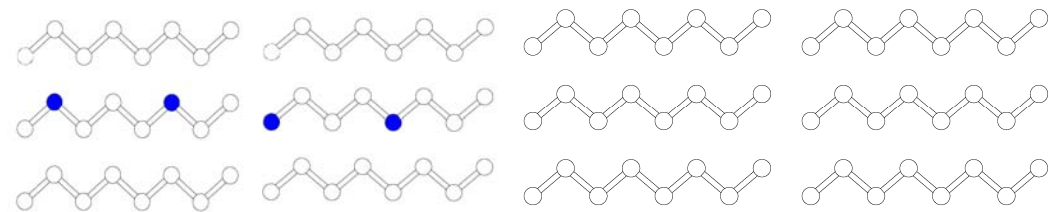

(d) A9

Figure 5.

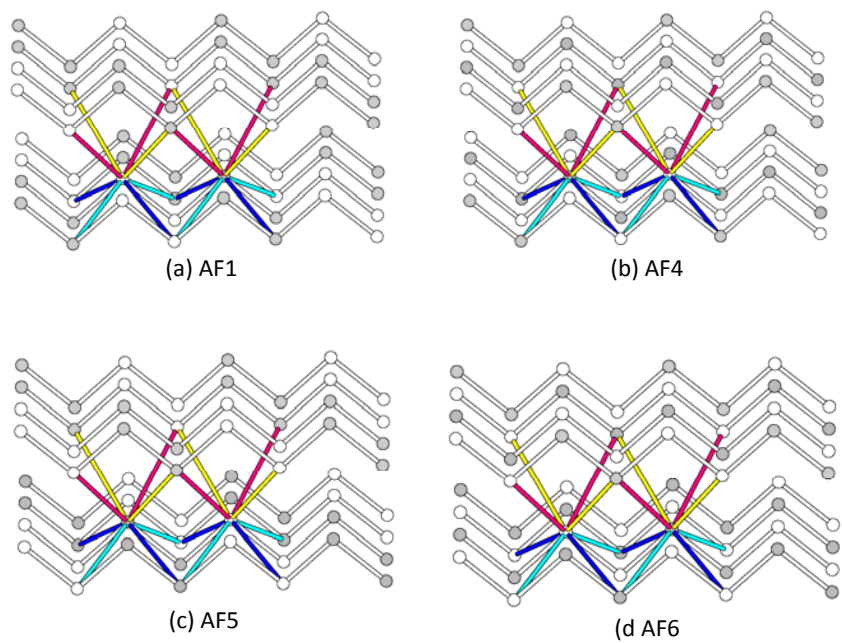

(a) AF1
(b) AF4
(c) AF5
(d) AF6

Figure 6.

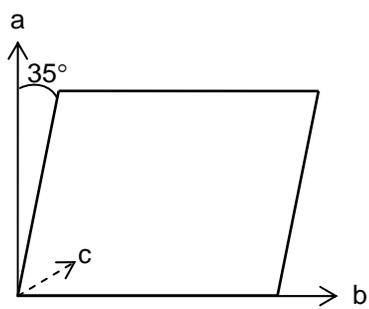

Figure 7.

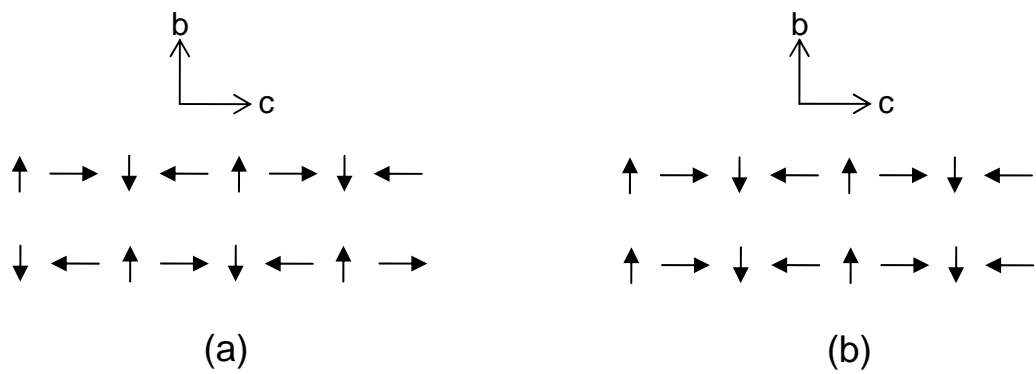

Figure 8.

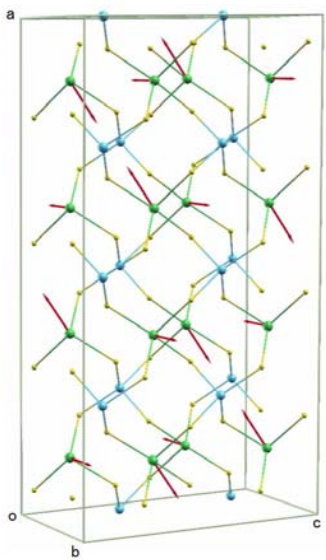

Figure 9.

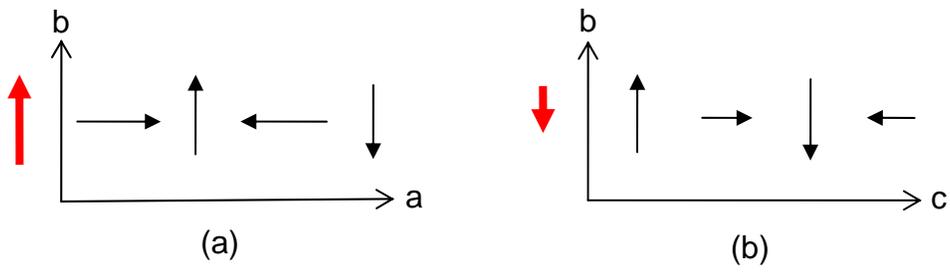

Figure 10.